\documentclass[pra,twocolumn,showpacs,preprintnumbers,amsmath,amssymb,ti
ghtenlines]{revtex4}

\usepackage{graphicx}
\usepackage{dcolumn}
\usepackage{bm}

\begin{document}

\preprint{\today}
\title{General limit to non-destructive optical detection of atoms}
\author{J.J. Hope and J.D. Close}
\affiliation{Australian Centre for Quantum Atom Optics,
Department of Physics,
Australian National University,
ACT 0200, Australia}

\email{joseph.hope@anu.edu.au}

\begin{abstract}
We demonstrate that there is a fundamental limit to the sensitivity of phase-based 
detection of atoms with light for a given maximum level of allowable spontaneous emission.
This is a generalisation of previous results for two-level and three-level atoms.
The limit is due to an upper bound on the phase shift that can be imparted on a laser 
beam for a given excited state population.  Specifially, we show that no 
single-pass optical technique using classical light, based on any number 
of lasers or coherences between any number of levels, can exceed the 
limit imposed by the two-level atom.
This puts significant restrictions on potential non-destructive optical measurement schemes.
\end{abstract}

\maketitle

\section{Introduction}
The prospect of continuously monitoring a cloud of cold atoms through a non-destructive, high-bandwidth measurement is very attractive.  It would be of particular interest for the study of quantum degenerate gases such as Bose-Einstein condensates (BEC).  Such a technique could be exploited to answer fundamental questions about the growth of condensates, spin dynamics in condensates, or vortex nucleation, turbulence and instabilities \cite{BECreview}. Alternatively, it could be used in applications such as the detection element in a feedback loop to force single-mode operation of an atom laser \cite{Simon}. The question is whether such a measurement is possible and, if so, how such a measurement should best be performed. Due to the compatibility of optical detection with cold atom experiments, nearly all experimental data on BECs  has been gathered using optical imaging. We concentrate on optical detection in this paper. 

Absorption imaging \cite{Ketterle1999}  and interferometric techniques such as phase contrast imaging \cite{Andrews}, spatial heterodyne imaging \cite{Walker} and polarisation contrast imaging \cite{Hulet} have all been applied to the study of cold atoms, and have all exploited the physics of the interaction of two level atoms with classical light fields. Although interferometric techniques exhibit an advantage in signal to noise ratio (SNR) in the minimally destructive measurement of the column density of an optically thick cloud, there is an upper bound to their sensitivity. In the shot noise limit, the SNR in an interferometric measurement is a function only of the measurement bandwidth and the destruction (spontaneous emission rate). In the limit of optically thin clouds, the SNR of absorption measurements and interferometric measurements under the same conditions of bandwidth and destruction are identical. In this way, there is a hard limit on the SNR achievable from any single pass technique based on the two level atom and classical light fields \cite{Lye03}.

Effects such as the enhancement of refractive index \cite{Scully} and the enhancement of dispersion close to a dark state that leads to phenomena such as slow or stopped light \cite{Hau} suggest strongly that it may be possible to beat the two level atom detection limit by manipulating coherences in a multi-level atom. It is known, for example, that in three-level systems
in the presence of a strong second laser, it is possible for a weak
probe beam to experience a non-zero phase shift without any absorption,
suggesting that manipulation of coherences in a three-level system
might provide a less destructive detection method
\cite{Arimondo,Scully}.  In terms of non-destructive optical detection however, the relevant question to ask is not whether a finite phase shift can exist without absorption but whether it can exist without excited state population, as it is the excited state population that drives spontaneous emission and heating of the condensate.  

The question of minimally destructive detection for the {\it restricted case of three level atoms} was addressed in our previous Letter \cite{Hope2005}. In that work, we found that in a minimally-destructive, optical measurement of the column density of cold atoms, no  single-pass technique exploiting classical light and coherences in a three level atom can beat the signal to noise limit imposed by the two-level atom.  The purpose of this paper is to examine all other possible schemes for generating large phase shifts for the purpose of non-destructive detection.  The question this paper addresses is: {\it Is it possible with an arbitrary atomic level scheme to produce a phase shift on a probe beam without a correspondingly high spontaneous emission rate?} Or more succinctly: {\it Can we beat the non-destructive detection limit imposed by the two level atom?} As we show in this paper, the answer to this more general question is still: No. 

For the three level atom, the proof of the theorem in the linear regime was relatively straightforward and proceeded by identifying the appropriate off diagonal element of the density matrix as the linear complex susceptibilty, up to a multiplicative constant. In the non-linear regime, even for a three level atom, it was unclear how to prove the  theorem using Maxwell's equations and an expansion of the polarization to all orders in the electric field. For an arbitrary number of  coherences between an arbitrary number of levels, as we  treat in this paper, the proof appears intractable using this approach.  To prove the results in our previous Letter and in the present paper, we developed a new technique for calculating phase shifts in both the linear and non-linear regimes by identifying the link between the phase shift on a light beam and the light shift of the dressed eigenstates of the full  Hamiltonian. In the present  paper, we have used this method to generalize our earlier proof  to any number of levels and coherences.

\section{Phase shift limits in multilevel systems}

We are searching for non-destructive optical detection of atoms.  The non-destructive criteria are:
\begin{enumerate}
  \item Atoms must not undergo spontaneous emission events.
  \item Atoms must not diffuse in momentum, and
  \item Atoms must return to their original electronic state after interaction with the light.
\end{enumerate}
Under these constraints, it is surprisingly easy to demonstrate that a phase shift on any laser beam interacting with the atoms must be associated with a particular minimum excited state population. 
 
Consider an atom with $N$ relevant electronic energy levels interacting with $M$ laser fields.  The interaction picture Hamiltonian for this system in the rotating wave approximation is
\begin{eqnarray}
H & = & \sum_{n=1}^N{\hbar \Delta_n |n\rangle\langle
n|}+\sum_{j=1}^M\sum_{l=1}^{m_j}(g_{j l} \hat{a}_j^\dag |L_{j
l}\rangle\langle U_{j l}| + \mbox{adj.})
\label{eq:Ham}
\end{eqnarray}
where $\hbar \Delta_n$ is the energy of atomic level $|n\rangle$, $\hat{a}_j$
is the annihilation operator for the optical mode of the
$j^{\mbox{th}}$ laser, $m_j$ is the number of transitions caused by the
$j^{\mbox{th}}$ laser, $|L_{j l}\rangle$ and $|U_{j l}\rangle$ are the
lower and upper atomic energy levels respectively of the $l^{\mbox{th}}$
transition of the $j^{\mbox{th}}$ laser, and $g_{j l}$ is its dipole coupling strength given by
\begin{equation}
g_{j l} = \sqrt{\frac{\hbar \omega_j}{2 \epsilon_0 V}} d_{U_{j l} L_{j l}}
\label{eq:gdef}
\end{equation}
where $d_{X Y}$ is the dipole moment between the electronic states $X$ and $Y$, $\omega_j$ is the angular frequency of the $j^{\mbox{th}}$ laser, and $V$ is the quantisation volume.  As our first condition assumes that any spontaneous emission must be negligible, this Hamiltonian will be sufficient to describe the entire evolution of the system.  

The second condition limits our choices of lasers and atomic levels, as most configurations of lasers and levels will lead to momentum diffusion due to the scattering of photons from one laser field to another.  Although it is not immediately obvious, this requirement is equivalent to demanding that the interaction return the atom to its original electronic state and the laser fields to their original number of photons.  Conceivably, a single, coherently applied momentum kick to the atoms could be acceptable for non-destructive detection, and this would mean that a redistribution of photons between the laser fields might be allowable.  However, ruling out any process by which the momentum might diffuse, combined with the third condition that the atom must return to its original state, means that we are left with a more restrictive constraint.  The combined system of atoms and lasers must therefore reduce to a series of manifolds in which each atomic
level is associated with a definite number of photons in each optical
mode, or the system could cycle back to the original electronic state while scattering photons between the lasers, leading to a spreading in the momentum wavefunction.  

\begin{figure}
\includegraphics[width=7.5cm]{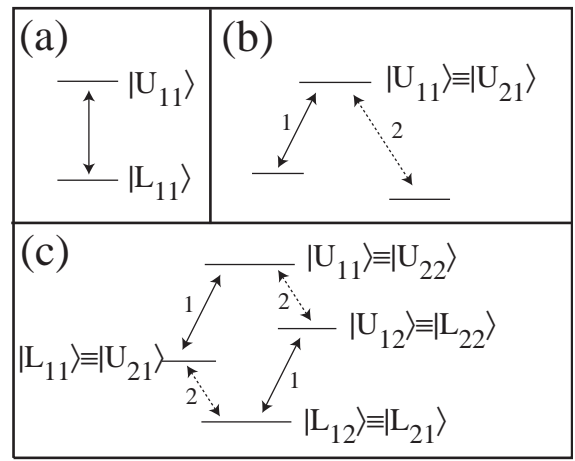}
\caption{Examples of lasers and atomic energy level schemes that are
allowable in our analysis.  (a) shows the two-level atom.  This can be
connected only by a single laser, or else it would be possible to
return to the ground state without returning to the original momentum
state.  (b) shows a Raman transition with two laser fields.  Each laser
still causes only one transition, but the excited state for each of
those transitions is identical, so in our notation $|U_{1
1}\rangle\equiv|U_{2 1}\rangle$.  (c) shows a more exotic system where
each laser causes multiple transitions but the atoms will still return
to the starting state without momentum diffusion.  In this example, all
four states are multiply defined in the $|U_{j l}\rangle, |L_{j
l}\rangle$ notation.}
\label{fig:levelexamples}
\end{figure}

For a two-level system, for example,
only one laser can couple the states $|1\rangle$ and $|2\rangle$.
During the interaction, the total quantum state will reduce to a set of
non-interacting manifolds of pairs of states $\{|1,n\rangle,
|2,n-1\rangle\}$, where $|j,n\rangle$ is the state with the atom in
electronic state $j$ with $n$ photons in the laser beam
\cite{Tannoudji}.  If more than one laser was used, Bragg scattering would occur.  A three-level system in a lambda configuration where
lower energy states $|1\rangle$ and $|3\rangle$ are each coupled to the
excited state $|2\rangle$ by a separate laser mode is another example of a system without coherent momentum diffusion.
In this case the independent manifolds are $\{|1,n,m\rangle,
|2,n-1,m\rangle, |3,n-1,m+1\rangle\}$.

In general, we can see that individual elements of the closed manifolds can be indexed by the atomic state
$|j\rangle$ alone, although the full description of the quantum state will include the number of
photons in each optical mode: 
\begin{equation}
\label{eq:stateshorthand}
|j\rangle \equiv |j,n_1+b_{1 j},n_2+b_{2 j}, \cdots,
n_M+b_{M j}\rangle.  
\end{equation}
In this notation $b_{i j}$ are the elements of an
integer-valued, constant matrix whose form is determined by our specific choices of lasers and levels.

In this reduced space, the Hamiltonian (\ref{eq:Ham}) is
\begin{eqnarray}
H & = & \sum_{n=1}^N{\hbar \Delta_n |n\rangle\langle
n|}+\sum_{j=1}^M\sum_{l=1}^{m_j}(g_{j l} \;\hat{\sigma}_{j l}^- \sqrt{n_j+1+b_{j L_{j l}}} + \mbox{adj.})
\label{eq:HamRed}
\end{eqnarray}
where 
\begin{eqnarray}
\hat{\sigma}_{j l}^- \equiv |L_{j l}\rangle \langle U_{j l}|,
\label{eq:sigmadef}
\end{eqnarray}
where we are using the shorthand notation defined in equation (\ref{eq:stateshorthand}).

This Hamiltonian can always be diagonalised, and although this will be a difficult procedure for complicated systems, we are fortunate not to need the explicit form of the eigenstates $|m(\{n\})\rangle$ or their eigenvalues $E_m(\{n\})$.  Note that these dressed states and dressed state energies may depend on the particular sub-manifold, and are hence indexed by the photon state $\{n\}$, which is shorthand for the set of number states $\{n_1,\cdots,n_M\}$.

The initial state of the total system before the interaction between the atoms and the light can be factorised into the initial atomic state $|I\rangle$, which may not necessarily be one of the electronic eigenstates, and the initial photonic state:
\begin{eqnarray}
|\psi_{\mbox{initial}}\rangle = \sum_{\{n\}} \alpha(\{n\})\;
|I, \{n\}\rangle.
\label{eq:photoninitial}
\end{eqnarray}
When the input laser fields are all coherent states, the coefficients $\alpha(\{n\})$ will factorise into the number basis expansion of a set of coherent state amplitudes describing each laser.

Each of the states $|I, \{n\}\rangle$ in the above expansion may be affected differently by the interaction between the light and the atoms.  This evolution is trivial in the dressed state basis, so we will expand these states in that basis, introducing the coefficients $c_m$:
\begin{eqnarray}
|I, \{n\}\rangle = \sum_{m=1}^M c_m(\{n\}) |m(\{n\})\rangle
\label{eq:eigenexpansion}
\end{eqnarray}
Just as the dressed states depend on the photonic state $\{n\}$, so do the coefficients of this expansion.  We are fortunate that due to our requirements for the behaviour of this system, we do not need to explicitly calculate either of these quantities.  Instead, we note that after the interaction, our second and third conditions require this state to return to $|I, \{n\}\rangle$ up to a phase factor.  Denoting the phase factor as $\varphi(\{n\})$ and noting that each dressed state will simply rotate in phase at a rate proportional to the dressed state eigenvalue, we find that
\begin{eqnarray}
\sum_{m=1}^M c_m(\{n\}) |m(\{n\})\rangle e^{-\frac{i E_m(\{n\}) \tau}{\hbar}} = |I, \{n\}\rangle e^{-i \varphi(\{n\})}
\label{eq:trivialevol}
\end{eqnarray}
where the interaction time is given by $\tau$.  Our final state is
\begin{eqnarray}
|\psi_{\mbox{final}}\rangle = \sum_{\{n\}} \alpha(\{n\})\;
|I, \{n\}\rangle e^{-i \varphi(\{n\})}.
\label{eq:photonfinal}
\end{eqnarray}

This final state will only correspond to a state with a phase shift $\Delta \phi_j$ on the $j^{\mbox{th}}$ laser if the total phase shift is linear in each photon number:
\begin{eqnarray}
\varphi(\{n\}) = \Delta \phi_1 n_1 + \cdots + \Delta \phi_M n_M,
\label{eq:purephaseshift}
\end{eqnarray}
but this will not always be the case.  In general, the final state could have much richer structure than a mere phase shift, but to the extent that each laser \textit{does} experience a phase shift, it will correspond to the partial derivative of the total phase with respect to the photon number of that laser.  The phase shift $\Delta \phi_j$ on the $j^{\mbox{th}}$ laser is therefore
\begin{eqnarray}
\Delta \phi_j = \frac{\partial \varphi(\{n\})}{\partial n_j},
\label{eq:actualphaseshift}
\end{eqnarray}
where, for systems where the phase shift is well-defined, the right hand side would be independent of $\{n\}$ in the region where $\alpha(\{n\})$ is non-negligible.  Where the phase shift is less well-defined, a good estimate would be obtained by evaluating the partial derivatives at the mean photon number for each optical mode.

Using (\ref{eq:eigenexpansion}), (\ref{eq:trivialevol}) and the orthogonality of the dressed states, we can see that
\begin{eqnarray}
e^{-\frac{i E_m(\{n\}) \tau}{\hbar}} = e^{-i \varphi(\{n\})} && \forall \;m: c_m\neq0
\label{eq:rotequiv}
\end{eqnarray}
which in combination with (\ref{eq:actualphaseshift}) leads directly to the relation
\begin{eqnarray}
\Delta \phi_j = \frac{\tau}{\hbar} \frac{\partial E_m(\{n\})}{\partial n_j} 
\label{eq:phaseshifttoE}
\end{eqnarray}
for all values of $m$ and $\{n\}$ where the coefficients $c_m(\{n\})$ and $\alpha(\{n\})$ are both non-trivial.

This relationship between the phase shift and the dressed state eigenvalues is at the heart of the theorem presented in this paper.  It was noted in \cite{Tannoudji} that such a relationship existed but that it was not a convenient method for calculating phase shifts.  The convenience of using it for our theorem is based on the fact that the relationship itself can prove our limit without referring to the details of any particular system.  The apparent redundancy in the $m$ index is not, as it first appears, a multiple definition of the phase shift.  If more than one dressed state is involved, then the system will exhibit transfer of population among the atomic energy levels during the interaction.  This will only satisfy the non-destructive criteria if the time of the interaction $\tau$ is precisely chosen such that condition (\ref{eq:rotequiv}) is true for all the relevant photon states, and therefore all the values of $m$ and $\{n\}$ corresponding to non-zero coefficients in the expansion.  This guarantees that equation (\ref{eq:phaseshifttoE}) is self-consistent.  A standard example of such a situation would be a two-level atom coupled by a near-resonant laser, where only certain interaction times will return the atom to the ground state.

Our previous work examined two and three-level atoms prepared in a single, dark dressed state, and showed that the relationship (\ref{eq:phaseshifttoE}) leads to a finite limit for the phase shift of each laser for a given excited state population of the atoms \cite{Hope2005}.  We are now in a position to extend that result to an arbitrary number of levels.  Rather than just considering atomic states prepared in a single dark state, the theory will allow the possibility of any non-trivial time-dependent dynamics, such as the multi-level generalisation of a $2 \pi$-pulse applied to a two-level atom.  In our formalism, this is equivalent to having non-zero population in multiple dressed states.  In order to recast our theorem to include these possibilities, a time-averaged population in the electronic state $|e\rangle$ is defined:
\begin{widetext}
\begin{eqnarray}
\bar{P}_e & = & \frac{1}{\tau} \int_0^\tau P_e(t) dt   =  \sum_{\{n\}} \left| \alpha(\{n\})\right|^2  \int_0^\tau \frac{dt}{\tau} \left|\sum_{m=1}^M c_m(\{n\}) \langle e|m(\{n\})\rangle e^{-\frac{i E_m(\{n\}) t}{\hbar}} \right|^2 
\nonumber \\
& = & \sum_{\{n\},m,m'} \left| \alpha(\{n\})\right|^2  c_m(\{n\}) c^*_{m'}(\{n\}) \langle e|m(\{n\})\rangle \langle m'(\{n\})|e\rangle \int_0^\tau \frac{dt}{\tau} e^{-\frac{i (E_m(\{n\})-E_{m'}(\{n\}) t}{\hbar}}
\label{eq:PebarTrans}
\end{eqnarray}
\end{widetext}
Under the constraint of equation (\ref{eq:rotequiv}), we can write
\begin{eqnarray}
e^{\frac{-i E_m(\{n\}) t}{\hbar}} & = & e^{i 2 \pi f_m(\{n\}) t/\tau} \; e^{-i \varphi(\{n\})}
\end{eqnarray}
where $f_m(\{n\})$ is an integer-valued function.  This means that the integral in equation (\ref{eq:PebarTrans}) is a Kronecker-delta function in $f_m(\{n\})$ and $f_{m'}(\{n\})$.  We can always be in the non-degenerate case by our initial choice of basis for any degenerate dressed states, so this can be reduced further to $\delta_{m, m'}$.  This leaves us with a very compact version of the average atomic population in the state $|e\rangle$:
\begin{eqnarray}
\bar{P}_e & = & \sum_{\{n\},m} \left| \alpha(\{n\})\right|^2  \left|c_m(\{n\})\right|^2 
\left|\langle e|m(\{n\})\rangle\right|^2
\label{eq:Pebar}
\end{eqnarray}

We can now return to our phase shift, and relate it to the average excited state populations in the atom.  From equation (\ref{eq:phaseshifttoE}) we can reformulate the derivatives with respect to the photon number to a more useful quantity:
\begin{eqnarray}
\Delta \phi_j = \frac{L}{\hbar c} \sum_{l=1}^{m_j} \frac{\partial E_m(\{n\})}{\partial \Omega_{j l}} \frac{\partial \Omega_{j l}}{\partial n_j}
\label{eq:phasetrans}
\end{eqnarray}
where we have introduced the length of the quantisation volume along the propagation axis of the $j^{\mbox{th}}$ laser field,  $L= c \tau$, and defined the functions $\Omega_{j l}$:
\begin{eqnarray}
\frac{\hbar \Omega_{j l}}{2} = g_{j l} \sqrt{n_j+1+b_{j L_{j l}}}
\end{eqnarray}
The factors of $\Omega_{j l}$ arise naturally in the Hamiltonian, and the notation is chosen to make a natural connection to the semiclassical picture.  Straightforward algebra can further transform equation (\ref{eq:phasetrans}) to involve standard optical quantities.  We find that
\begin{eqnarray}
\Delta \phi_j = \sum_{l=1}^{m_j} \frac{\partial E_m(\{n\})}{\partial \Omega_{j l}} 
\frac{\sigma_j \;\gamma_{j l} \; L}{2 \;\hbar \;\Omega_{j l} \;V}
\label{eq:phasetrans2}
\end{eqnarray}
where $k_j$ and $\sigma_j = \frac{6 \pi}{k_j^2}$ are the wavenumber and the single atom cross-section respectively of the $j^{\mbox{th}}$ laser, and $\gamma_{j l} = \frac{k_j^3 d_{U_{j l} L_{j l}}^2}{3 \pi \epsilon_0 \hbar}$ is the spontaneous emission rate from the upper state of its $l^{\mbox{th}}$ transition.  

Derivatives of the dressed state eigenvalues can be determined via the Hellman-Feynman theorem
\begin{eqnarray}
\frac{\partial E_m(\{n\})}{\partial \Omega_{j l}}  & = & \langle m(\{n\}) | \frac{\partial \hat{H}}{\partial \Omega_{j l}} | m(\{n\}) \rangle \nonumber \\
 & = & \frac{\hbar}{2}  \langle m(\{n\}) | \left(\sigma_{j l}^- +\sigma_{j l}^+\right) | m(\{n\}) \rangle \nonumber \\
 & = & \hbar  \Re\left\{\rho_{L_{j l} U_{j l}}^{(m,\{n\})}\right\}
\label {eq:HFtheorem}
\end{eqnarray}
where $\rho^{(m,\{n\})}=|m(\{n\})\rangle\langle m(\{n\})|$ is the density matrix describing the eigenstate.  The magnitude of the phase shift per atom can therefore be written
\begin{eqnarray}
\left|\Delta \phi_j\right| &=& \left|\sum_{l=1}^{m_j} \frac{\sigma_j \;\gamma_{j l} \; L}
{2 \;\Omega_{j l} \;V} \Re\left\{\rho_{L_{j l} U_{j l}}^{(m,\{n\})}\right\}\right|
\nonumber \\
&\leq&\sum_{l=1}^{m_j} \frac{\sigma_j \;\gamma_{j l} \; L}
{2 \;\Omega_{j l} \;V}  \left|\Re\left\{\rho_{L_{j l} U_{j l}}^{(m,\{n\})}\right\}\right|
\nonumber \\
&\leq&\sum_{l=1}^{m_j} \frac{\sigma_j \;\gamma_{j l} \; L}
{2 \;\Omega_{j l} \;V}  \left|\langle U_{j l}|m(\{n\})\rangle\right|
\label{eq:phasetrans3}
\end{eqnarray}
where the last line comes from the properties of density matrices that $|\rho_{L U}|^2\le \rho_{L L} \rho_{U U}$, and $\rho_{L L}\le1$.

This inequality holds for each eigenstate where $c_m(\{n\})$ and $\alpha(\{n\})$ are non-negligible. With the normalisation condition,
\begin{equation}\sum_{m,\{n\}}\left|\alpha(\{n\}\right|^2 \left|c_m(\{n\})\right|^2=1,
\label{eq:norm}
\end{equation}
this allows us to replace the right hand side of the inequality with a sum:
\begin{widetext}
\begin{eqnarray}
\left|\Delta \phi_j\right| &\leq& \frac{L\;\sigma_j}{2 \;V} \sum_{l=1}^{m_j} \frac{\gamma_{j l}}
{\Omega_{j l}} \sum_{m,\{n\}}\left|\alpha(\{n\}\right|^2 \left|c_m(\{n\})\right|^2
\left|\langle U_{j l}|m(\{n\})\rangle\right|
\nonumber \\
&\leq& \frac{L\;\sigma_j}{2 \;V} \sum_{l=1}^{m_j} \frac{\gamma_{j l}}
{\Omega_{j l}} 
\sqrt{\sum_{m,\{n\}}\left|\alpha(\{n\})\right|^2 \left|c_m(\{n\})\right|^2}
\sqrt{\sum_{m,\{n\}}\left|\alpha(\{n\})\right|^2 \left|c_m(\{n\})\right|^2
\left|\langle U_{j l}|m(\{n\})\rangle\right|^2}
\nonumber \\
&=& \frac{L\;\sigma_j}{2 \;V} \sum_{l=1}^{m_j} \frac{\gamma_{j l}}
{\Omega_{j l}} 
\sqrt{\sum_{m,\{n\}}\left|\alpha(\{n\})\right|^2 \left|c_m(\{n\})\right|^2
\left|\langle U_{j l}|m(\{n\})\rangle\right|^2}
\label{eq:phasetrans4}
\end{eqnarray}
\end{widetext}
where the second line is obtained by the Cauchy-Schwartz inequality, and the last line is recovered by the normalisation condition (\ref{eq:norm}).  From equation (\ref{eq:Pebar}), we can now write a bound on the phase shift on the laser beam due to a single atom in terms of the excited state populations:
\begin{eqnarray}
\left|\Delta \phi_j\right|\bigg|_{\mbox{per atom}} &\leq& \frac{L\;\sigma_j}{2 \;V} \sum_{l=1}^{m_j} \frac{\gamma_{j l}}
{\Omega_{j l}} \sqrt{\bar{P}_{U_{j l}}}
\label{eq:phasePerAtom}
\end{eqnarray}
To account for the effect of multiple atoms within the quantisation volume, we add the individual phase shifts.  Our final result depends on the column density $\tilde{n}=\rho L$ of the atoms:
\begin{eqnarray}
\left|\Delta \phi_j\right|\bigg|_{\mbox{total}} &\leq& \frac{\tilde{n}\;\sigma_j}{2} \sum_{l=1}^{m_j} \frac{\gamma_{j l}}
{\Omega_{j l}} \sqrt{\bar{P}_{U_{j l}}}
\label{eq:phaseTotal}
\end{eqnarray}

It appears from this result that the phase shift on each field can be made large for a given excited state population, provided the field strength, and hence the Rabi frequency $\Omega_j$, becomes very small.  The detection of the phase shift of an arbitrarily weak signal is arbitrarily difficult, however, and a true measure of the sensitivity of an atomic density measurement is the signal to noise ratio.  This will ultimately be limited by the shot noise.

The signal to noise ratio (SNR) of a shot-noise limited measurement of the phase of the $j^{\mbox{th}}$ laser by an ideal, single-pass interferometer based on a coherent local oscillator is
\begin{equation}
\label{eq:idealSNR}
SNR_j = \sqrt{\frac{\eta\;P}{B\; \hbar \;\omega_j}} \left|\Delta \phi_j\right|
\end{equation}
where $\eta$ is the quantum efficiency of the photodetectors, $P$ and $\omega_j$ are the power and angular frequency respectively of the laser, and $B$ is the temporal bandwidth of the measurement.  Combining this with equation (\ref{eq:phaseTotal}) gives us a fundamental limit for the SNR:
\begin{eqnarray}
SNR_j &\leq& \sqrt{\frac{\eta\;A\;c\;\epsilon_0\;E_j^2}{2\;B\; \hbar \;\omega_j}}
\frac{\tilde{n}\;\sigma_j}{2} \sum_{l=1}^{m_j} \frac{\gamma_{j l}}
{\Omega_{j l}} \sqrt{\bar{P}_{U_{j l}}}
\label{eq:SNRtranslimit}
\end{eqnarray}
where $A$ is the cross-sectional area of the beam and $E_j$ is its electric field.  After some algebra, this leads to
\begin{eqnarray}
SNR_j &\leq& \sum_{l=1}^{m_j} \frac{\tilde{n}}{2}\sqrt{\frac{\eta\;A\;\sigma_j\;\bar{P}_{U_{j l}}\;\gamma_{j l}}{B}}
\label{eq:SNRlimit}
\end{eqnarray}
Alternatively, this limit can be written in terms of the average spontaneous emission rates $\Gamma_{j l} = \bar{P}_{U_{j l}} \gamma_{j l}$:
\begin{eqnarray}
SNR_j &\leq& \sum_{l=1}^{m_j} \frac{\tilde{n}}{2}\sqrt{\frac{\eta\;A\;\sigma_j\;\Gamma_{j l}}{B}}
\label{eq:SNRlimit2}
\end{eqnarray}
This is the main result of this paper, showing that SNR for any phase-based measurement of the atomic density using classical states is bounded above by a limit that depends only on the temporal and spatial bandwidth, fixed atomic parameters, and the spontaneous emission rate.  This is true for any number of lasers combined with any level scheme.  

The detection limit shown here includes, and is essentially the same as, that derived for the two-level atom interacting with a single laser.  This simple scheme is the fundamental basis for all current attempts at non-destructive atom detection, and our result shows that there is no advantage in using a more complicated process to produce the phase shift.

\section{Conclusions}

In this paper we have proved that for a minimally-destructive, phase-based optical measurement of the column density of cold atoms, no single-pass technique exploiting classical light and {\it any number of coherences between any number of levels}, can exceed the signal to noise limit imposed by the two-level atom. This places significant restrictions on the design of measurement schemes for the detection of cold atoms.  The limit on the phase shift for a given excited state population in equation (\ref{eq:phaseTotal}) holds for any non-destructive detection candidate. The limit to the measurement of that phase in equation (\ref{eq:idealSNR}), and therefore the detection limit as described in (\ref{eq:SNRlimit2}), is specific to classical states of light, and for single-pass phase measurement techniques.  The complete list of candidates for superior non-destructive atomic measurement is therefore:

\begin{enumerate}
  \item Detection via non-optical methods.
  \item Optical detection that does not rely on absorption or phase shift measurements.
  \item Optical phase shift measurement using non-classical (e.g. squeezed) light.
  \item Optical phase shift measurement using multi-pass interferometry.
\end{enumerate}

The first two items on the list are obvious avenues for exploration, although it might be noted that so far in the field of experimental atom optics, the only atomic detection not based on the absorption or phase-shift of light has required a collision with the detector.  These methods clearly can not be regarded as candidates for non-destructive measurement, so completely new techniques would have to be developed.

Squeezing slightly in excess of 6 dB at 1064 nm has been achieved through optical parametric oscillation in crystals or in fibres through the Kerr effect  \cite{Lam, Breitenbach, Silberhorn}. If similar squeezing were produced at one of the wavelengths commonly used for BEC research, this would represent a maximum improvement of a factor of 4 in the SNR of a measurement on cold atoms. Although it is possible to produce squeezed light at these wavelengths in principle, the asymptotic fragility of strong squeezing suggests that this method will never produce significant gains in sensitivity.  
 
Using resonant interferometry provides a factor of the square root of the finesse of the system \cite{Lye03, Hinds}. This appears to be a far more likely avenue. Cavities with finesses approaching or in some cases exceeding $10^5$ have been demonstrated in cavity QED experiments with single atoms \cite{Rempe, Kimble}, suggesting an improvement of up to three orders of magnitude in the SNR in the measurement of condensates.  It is a daunting prospect  to combine this technology with BEC but  it appears that it would lead to significant improvement in the measurement of these systems. Although there are many practical questions yet to be addressed in such a measurement, the direction does appear promising.  
 
\begin{acknowledgments}
The Australian Centre for Quantum Atom Optics is an Australian Research
Council Centre for Excellence. 
\end{acknowledgments}

\end{document}